\documentclass{nature}
\usepackage[per-mode=symbol]{siunitx}
\usepackage{graphicx}
\usepackage[symbol]{footmisc}
\usepackage[dvipsnames]{xcolor}

\usepackage{soul}
\sethlcolor{White}

\title{Origin of strong-field induced low-order harmonic generation in amorphous solids}
\author{P. J{\"u}rgens$^{1,*}$, B. Liewehr$^{2,*}$, B. Kruse$^{2,}$\footnote[1]{P.J., B.L. \& B.K. contributed equally to this work} , C. Peltz$^{2}$, D. Engel$^{1}$, A. Husakou$^{1}$, T. Witting$^{1}$, M. Ivanov$^{1}$, M. J. J. Vrakking$^{1}$, T. Fennel$^{1,2}$ \& A. Mermillod-Blondin$^{1}$}
\begin{document}
\maketitle
\begin{affiliations}
 \item Max-Born-Institute for Nonlinear Optics and Short Pulse Spectroscopy, Max-Born-Str. 2A, D-12489, Berlin, Germany
 \item 	Institute of Physics, University of Rostock, Albert-Einstein-Str. 23, D-18059, Rostock, Germany
\end{affiliations}

\begin{abstract}
%
Kerr-type nonlinearities form the basis for our physical understanding of nonlinear optical phenomena in condensed matter, such as self-focusing, solitary waves, and wave mixing~\cite{Boyd2002, Shen2002,Agrawal2013}.
In strong fields, they are complemented by higher-order nonlinearities that enable high harmonic generation, which is currently understood as the interplay of light-driven intraband charge dynamics and interband recombination~\cite{Vampa2015,Schubert2014,Ghimire2011}. Remarkably, the nonlinear response emerging from the associated sub-cycle injection dynamics of electrons into the conduction band has been almost completely overlooked in solids and only partially considered in the gas phase~\cite{Mitrofanov2011,Brunel1990,Siders2001,Verhoef2010}. Here we reveal this strong-field-induced nonlinearity in amorphous wide-bandgap dielectrics by means of time-resolved, low-order wave mixing experiments and show that close to the material damage threshold
the so far unexplored injection current provides the leading contribution. The sensitivity of the harmonic emission to the sub-cycle ionization dynamics offers an original approach to characterize the evolution of laser-induced plasma formation in optical microprocessing.
\end{abstract}

When solid dielectrics are exposed to an intense infrared laser field, the quasi-instantaneous interband polarization associated with the Kerr effect~\cite{Boyd2002} is accompanied by electron injection into the conduction band near the peaks of the oscillating laser field via strong-field ionization~\cite{Krausz2009,sommer2016} (step 1 in Fig.~1).
After subsequent laser-driven acceleration of the electron in the conduction band (step 2), interband recombination with the hole left behind (step 3) can lead to the emission of high-harmonics of the driving field \cite{Vampa2015a, Wang2017}, in analogy to high-harmonic-generation (HHG) in the gas phase~\cite{Corkum1993, Lewenstein1994}. The non-parabolic landscape of the valence and conduction bands along which electrons and holes are driven by the laser field in step 2 can give rise to an additional intraband contribution to the nonlinear response that has no equivalent in atomic HHG~\cite{Schubert2014, Ghimire2011, Luu2015}. The sensitivity of the inter- and intraband mechanisms to the band structure and crystal orientation with respect to the laser polarization has turned high harmonic spectroscopy into an important new tool for optical characterization of the electronic structure and dynamics of solids~\cite{Vampa2015,Vampa2017, Lanin2017, Luu2015}.

\begin{figure}
\centering
\includegraphics[width=0.5\textwidth]{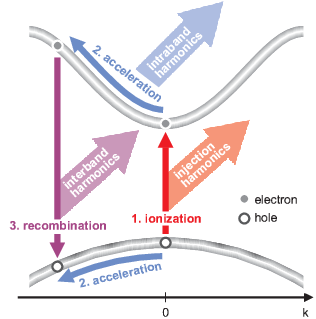}
\caption{
\textbf{Schematic description of harmonic generation in solids.} The strong-field induced electron dynamics can be divided into three major phases. It begins with interband excitation of electrons from the valence to the conduction band via strong-field ionization (step 1) followed by laser field-driven intraband motion of the excited electron in the conduction band and the remaining hole in the valence band (step 2), and can finally lead to interband recombination of the electron with the hole (step 3). Each of these steps is associated with distinct nonlinearities that contribute to the harmonic emission (as indicated). Our work demonstrates the dominant contribution of the injection step for the emission of low-order harmonics from amorphous wide-bandgap dielectrics for laser intensities close to the damage threshold.}
\label{fig:intro}
\end{figure}

Besides contributions from intraband currents and interband recombination that dominate HHG in dielectric solids, the sub-cycle dynamics of the injection of valence electrons into the conduction band in step 1 can lead to harmonics as well, as suggested by Brunel~\cite{Brunel1990} in the context of HHG in the gas phase. Brunel considered a stepwise injection of electrons into the continuum via tunnel ionization, with one step at the peak of each laser half cycle, and studied the low-order harmonic emission associated with the interaction of this temporally modulated electron population with the laser field. Despite the well-known impact of sub-cycle ionization dynamics on the absorption of intense laser pulses in gases~\cite{Geissler1999} and solids~\cite{Yabana2012,sommer2016}, the role of injection-induced nonlinearities in the wave mixing inside solids remains essentially unknown, though a pioneering experiment provided indications for their significance~\cite{Mitrofanov2011}.
Of particular interest is the relative impact of such nonlinearities compared to all other contributions including the Kerr response. Here we present time-resolved, two-color wave mixing experiments that provide evidence for the dominance of injection-induced low-order harmonic emission at intensities near the material damage threshold. Most importantly, our results expose the underlying mechanism and show that the corresponding nonlinear response does not reflect the conventional Brunel contribution but stems from the so far overlooked contribution of the injection current. The resulting nonlinear polarization is associated with the charge displacement resulting from the strong-field induced interband excitation process itself, in analogy to the current resulting from tunneling in a low-frequency laser field~\cite{Geissler1999}.

\begin{figure}
\centering
\includegraphics[width=0.9\textwidth]{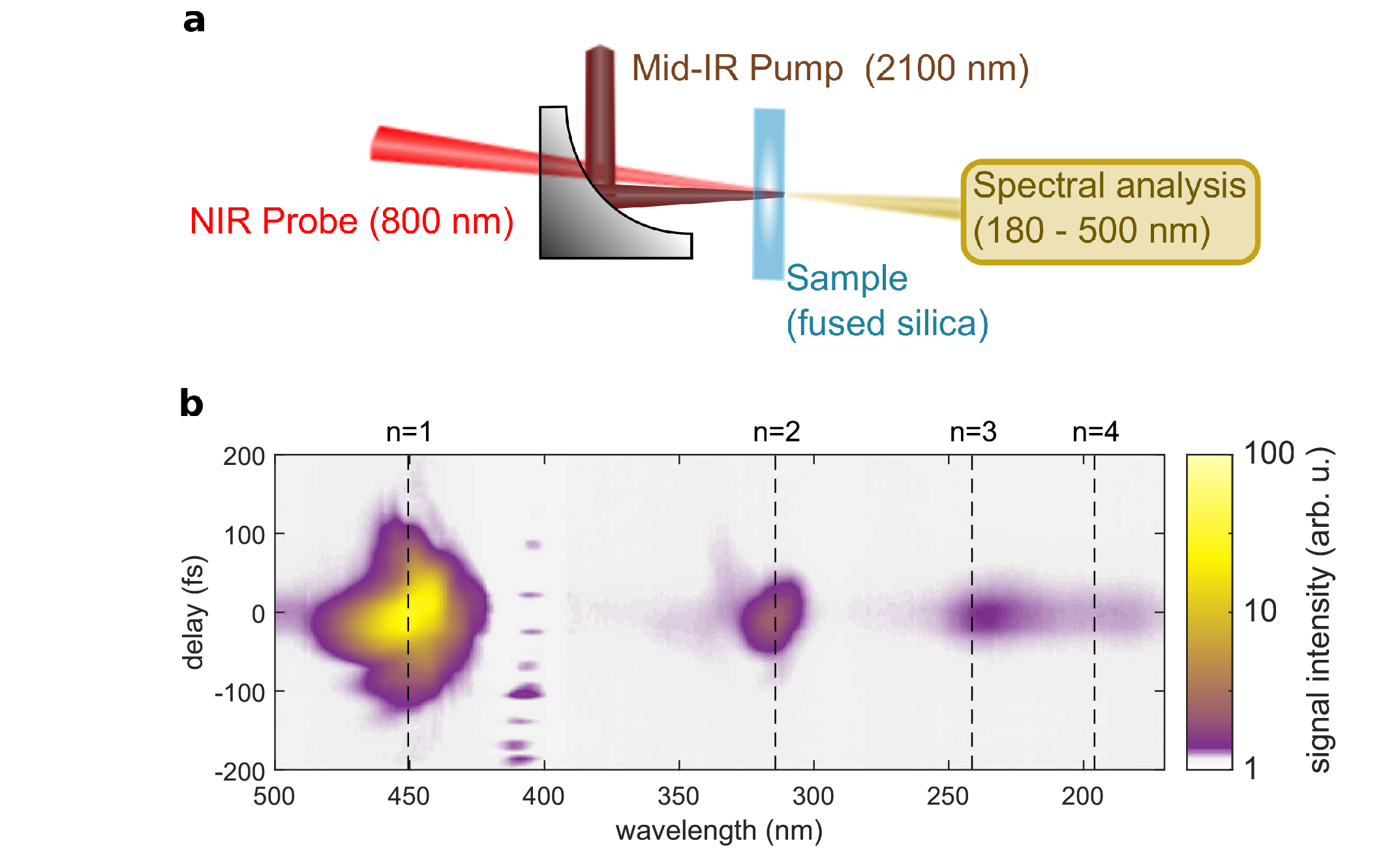}
\caption{\textbf{Time-resolved measurement of low-order two-color harmonics.} (a) Schematics of the experimental setup. An intense mid-IR pump laser pulse is overlapped with a weak, time-delayed NIR probe laser pulse in the bulk of a 0.5\,mm thick UV-grade fused silica sample. The generated low-order harmonic radiation is analyzed using a UV/VIS spectrometer. (b) Spectrally resolved low-order harmonic emission from the fused silica sample measured as a function of the delay between a strong pump pulse with 2.1~$\mu$m central wavelength focused to an intensity of 12 TW/cm$^2$ and a weak probe pulse at 0.8~$\mu$m central wavelength focused to an intensity of 0.015 TW/cm$^2$. The pump and probe pulses were provided with parallel polarizations and have durations of 140 fs and 45 fs, respectively. We verified that only a single probe photon participated in the underlying  wave mixing processes by demonstrating the linear dependence of all observed signals on the intensity of the probe beam (see Fig. S2 in the Supplementary Information). The dashed lines indicate the expected central wavelengths of the two-color harmonics according to Eq.~(\ref{eq:harmonic_orders}). Note that similar spectrograms were also recorded for perpendicularly polarized pulses.}
\label{fig:exp1}
\end{figure}

In the experiment, an intense mid-IR pump laser pulse ($\lambda_{\rm pump}=2.1\,\mu$m) and a weak, time-delayed NIR probe laser pulse ($\lambda_{\rm probe}=800\,$nm) were focused into a thin fused silica plate (\SI{0.5}{\milli\meter} thickness, \SI{7.7}{\electronvolt} bandgap) using a close-to-collinear pump-probe geometry (Fig.~\ref{fig:exp1}a). The resulting low-order harmonic emission in the visible and ultra-violet wavelength range was recorded as a function of the pump-probe delay.
In the region of pump-probe overlap, intense harmonic emission was observed at frequencies described by
\begin{equation}\label{eq:harmonic_orders}
\omega_n = 2\,n\, \omega_{\textrm{pump}} + \omega_{\textrm{probe}},
\end{equation}
where $n = 1, 2, 3, 4$ characterizes the order of the wave mixing (see Fig~\ref{fig:exp1}b).
As all harmonics lie well below the bandgap, we immediately rule out interband recombination as a possible source of the detected radiation.

\begin{figure}
\centering
\includegraphics[width=0.8\textwidth]{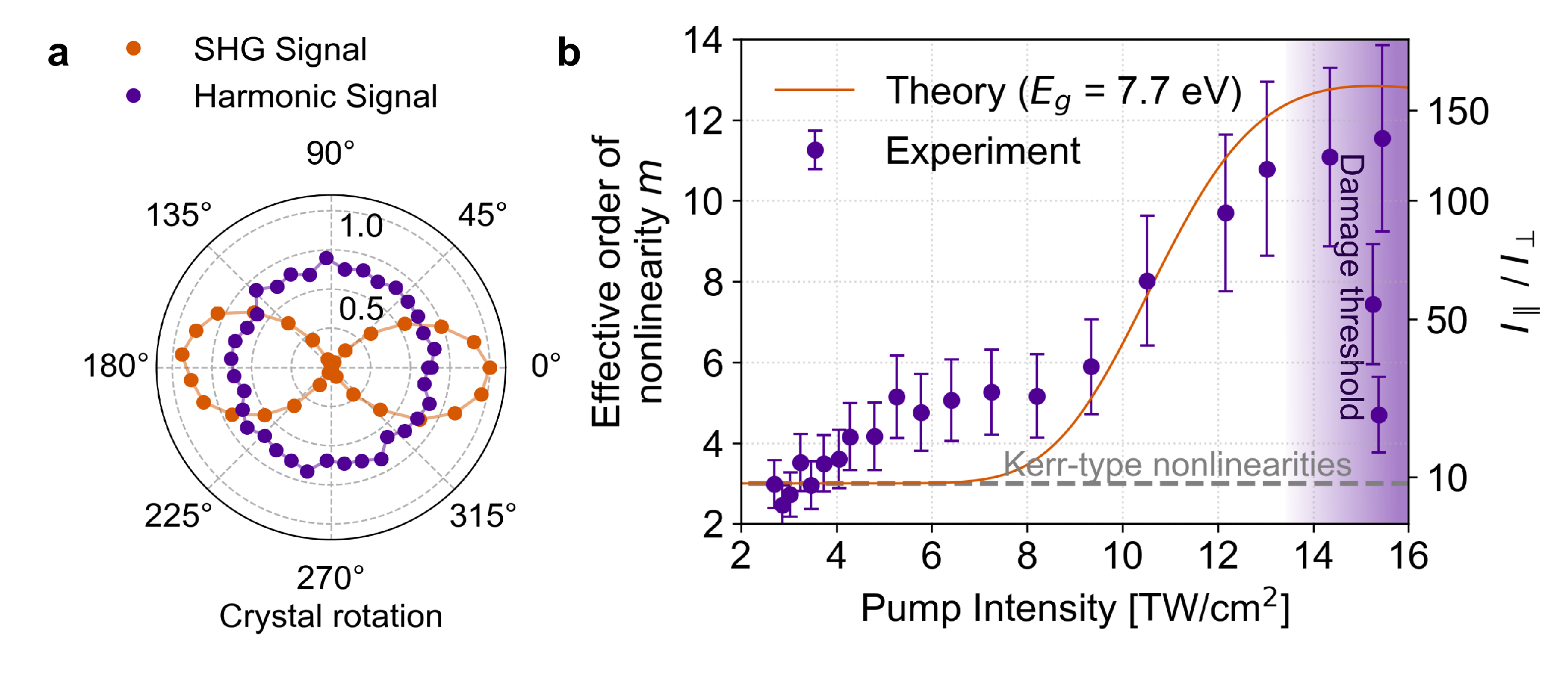}
\caption{\textbf{Orientation and intensity dependence of the harmonic emission.} {\bf(a)} Dependence of the $n=1$ harmonic emission yield from crystalline SiO$_2$ on the angle between the principal axis of the crystal and the polarization axis of the pump laser pulse (purple), measured at a pump laser intensity of $\SI{11.5}{\tera\watt\per\square\centi\meter}$ and for perpendicular polarization of pump and probe pulses. The probe pulse intensity was $I_{\textrm{probe}}\approx$~\SI{0.05}{\tera\watt\per\square\centi\meter}.
The angular dependence of second harmonic generation (SHG) of the probe is shown as well (orange) and provides a reference orientation. Whereas the SHG signal depends on the crystal orientation, the measured two-color harmonics do not, ruling out carrier oscillations in the anharmonic valence or conduction bands as their origin. {\bf (b)} Effective order of nonlinearity as extracted from the ratio of the $n=1$ harmonic yield obtained with parallel ($I_{\parallel}$) and perpendicular polarization ($I_{\perp}$) of the pump and probe beams as a function of the pump pulse intensity. The dashed line represents the result expected for a Kerr-type nonlinearity with $m=3$ alone and the orange solid curve shows the effective order of nonlinearity predicted by our numerical model for a bandgap of 7.7\,eV. Note that the right scale displays the associated intensity ratio. }
\label{fig:exp2}
\end{figure}

Additional experiments were performed to distinguish the contributions of all other remaining mechanisms to the nonlinear response. The formation of harmonics from intraband currents relies on the anharmonicity of the valence and conduction bands of the sample including their structural anisotropies~\cite{You2016}. To analyze corresponding crystal orientation effects, the experiments were repeated in Y-cut crystalline quartz samples. From the observation that the two-color harmonic emission from the crystalline target shows no
 pronounced orientation dependence (Fig.~\ref{fig:exp2}a), as in the case of the amorphous sample, we conclude that the observed harmonics do not arise from field-driven intraband currents.

Next, to explore the possible origin of the observed harmonics in Kerr-type nonlinear polarization we developed an approach to estimate the effective order of the underlying nonlinearity. We consider an effective nonlinear polarization
\begin{equation}\label{eq:nlo}
\mathbf{P} = \varepsilon_0 \chi^{(2\mu + 1)} \vert \mathbf{E} \cdot \mathbf{E} \vert ^{\mu} \cdot \mathbf{E}
\end{equation}
that follows the instantaneous electric field $\bf{E}$ according to a power law with an effective order $m=2\mu+1$ and an associated effective nonlinear susceptibility $\chi^{(m)}$. As derived in the Supplementary Information, within this model the nonlinearity can be characterized without measuring the intensity dependence of the harmonic emission. Instead, the effective order $m$ is fully determined by the ratio of the yield of the two-color harmonic emission observed with parallel ($I_{\parallel}$) and perpendicular ($I_{\perp}$) pump and probe polarizations. The relation $m^2=I_{\parallel} / I_{\perp}$ follows for wave mixing processes that involve, as realized in our experiment, only a single probe photon.

The measured ratio $I_{\parallel } / I_{\perp}$  reveals a swift growth of the effective order with pump intensity (see Fig. 3b). Departing from an order corresponding to a Kerr response with $m=3$ that increases and clearly saturates near $m=5$ for intensities up to $\approx$~\SI{8}{\tera\watt\per\square\centi\meter}, the effective order grows rapidly and reaches orders $m>10$ at intensities close to the damage threshold. Correspondingly, we conclude that the contribution of the Kerr-nonlinearities to the signals shown in Fig.~\ref{fig:exp1}b is marginal.



As interband, intraband and Kerr-type nonlinearities have been excluded as dominant sources of the harmonics, we finally examine whether an injection-driven nonlinearity can explain the observations. We model the harmonic signal from both the sub-cycle dynamics of electron transfer to the conduction band via strong-field interband excitation and the subsequent acceleration, neglecting effective mass effects and assuming a parabolic conduction band to eliminate contributions from band anharmonicities and Bloch oscillations due to reflections at zone boundaries. The resulting emitted field is proportional to the time-derivative of the associated current density
\begin{equation}\label{eq:harm}
\dot {\mathbf{J}\,} = q_e n_0 \left[ \frac{q_e}{m_e} \mathbf{E} \rho + \mathbf{v_0} \dot{\rho} + \frac{\partial}{\partial t}(\mathbf{x_0} \dot{\rho}) \right],
\end{equation}
where $\mathbf{E}$ is the driving laser field, $q_e=-e$ is the electron charge, $m_e$ the mass. Further, $\rho$ is the density of electrons in the conduction band (normalized to the molecular density $n_0$), which depends on the sub-cycle injection rate according to $\dot{\rho} = \Gamma(|\mathbf{E}|)$, and $\mathbf{v_0}$ and $\mathbf{x_0}$ are the birth velocity and the spatial displacement of the electrons after being promoted from the valence into the conduction band.

The current considered in Eq.~(\ref{eq:harm}) has been discussed in the context of strong-field induced absorption~\cite{Geissler1999} and contains a sum of three contributions. The first term describes the acceleration of conduction band electrons by the laser field as described by Brunel and creates a nonlinearity only if the electron density is rapidly modulated by ionization~\cite{Brunel1990}.
The second term describes a current due to the appearance of electrons in the conduction band with finite initial velocity and is negligibly small in our experiments due to the low photon energy of the pump beam. The third term describes a time-dependent injection current that is associated with the promotion of the electrons from the valence to the conduction band itself. In a quasistatic tunneling picture, this third term can be interpreted as the current through the potential barrier resulting from the combination of the Coulomb and laser fields. This current is also fundamental to maintain energy conservation, as the energy needed to realize the interband excitation process must appear as a loss in the field energy density $u$ through $\dot u= - \mathbf{J} \cdot \mathbf{E}$. In optical tunneling, energy conservation is fulfilled when this current reflects a spatial displacement of the electrons by $\mathbf{x}_0=\mathbf{E} \,E_{\rm g} /(q_e\,E^2)$, i.e. a displacement due to electron transport to the classical tunnel exit~\cite{Geissler1999}. Despite its relevance to energy conservation, this injection current $\mathbf{J}=\mathbf{x}_0 \dot \rho$ has so far not been considered as a source of nonlinearity leading to harmonic emission.

To analyze the impact of the injection-induced nonlinearities described by Eq.~(\ref{eq:harm}) we simulate the emitted field for the experimental pulse parameters using an instantaneous Ammosov-Delone-Krainov (ADK) tunneling rate~\cite{Ammosov_JETP_1986} and including the acceleration term ${\dot \mathbf{J}}_{\rm Kerr}= {\ddot  \mathbf{P}}_{\rm Kerr}$, where ${\bf P}_{\rm Kerr}$ is the conventional nonlinear Kerr polarization for SiO$_2$ as described by Eq.~(\ref{eq:nlo}) using $m=3$ (see Supplementary Information).
To avoid ambiguity, no empirical terms beyond the well-known $\chi^{(3)}$ response for SiO$_2$ were included. The predicted effective order of the two-color harmonic emission is determined as in the experiment from the ratio $I_{||}/I_{\perp}$ and reproduces the experimentally observed rapid growth for intensities beyond 8\,TWcm$^{-2}$ in Fig.~\ref{fig:exp2}b. We thus conclude that the injection-current induced nonlinearities are the source of the wave mixing signal observed in the high intensity range. We emphasize that our simplified model does not contain free adjustable parameters.

\begin{figure}
\centering
\includegraphics[width=0.6\textwidth]{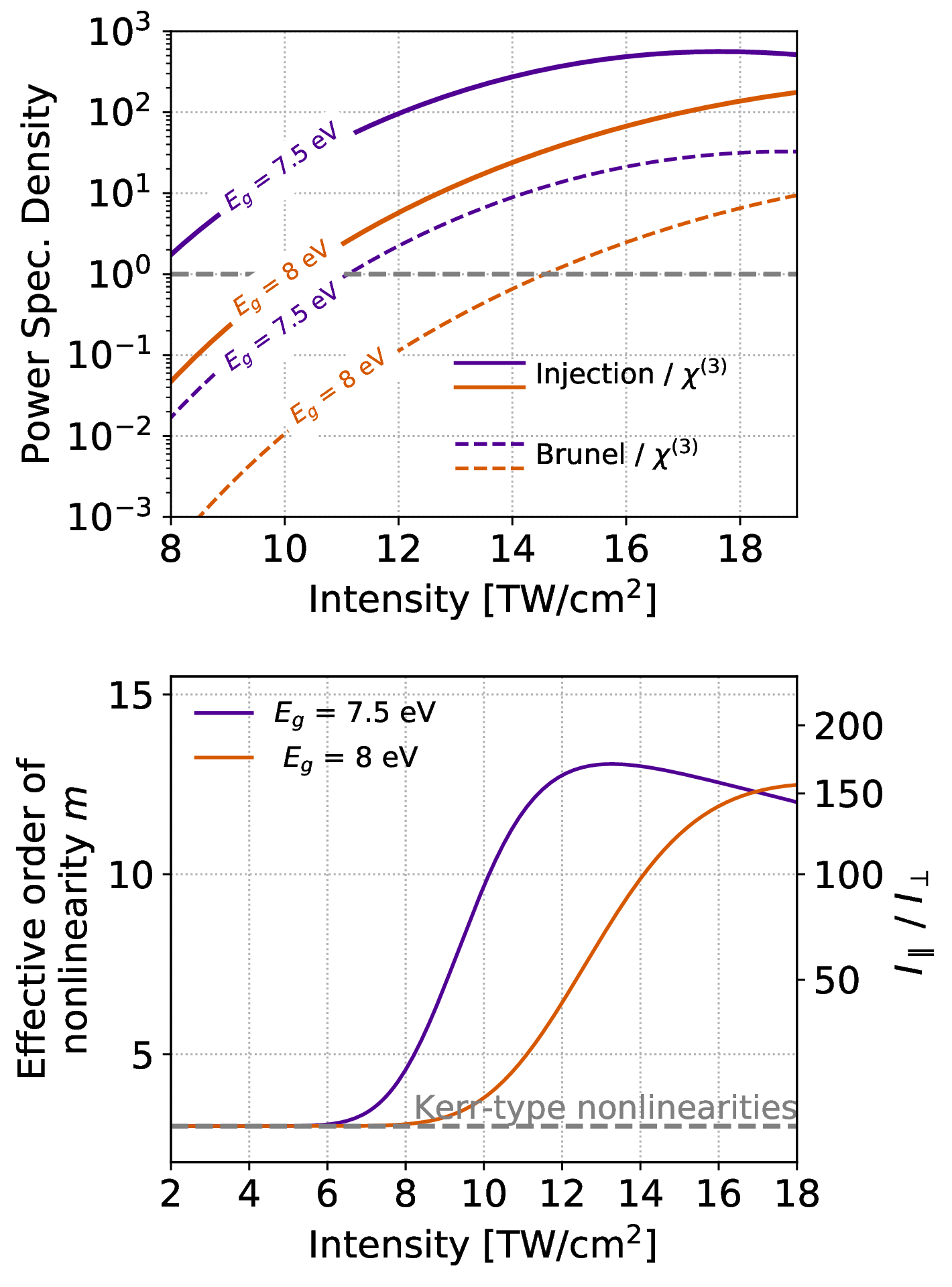}
\caption{\textbf{Comparison of the predicted contributions to the nonlinear response.} {\bf (a)} Relative magnitude of the contributions of the injection current and the Brunel term to the formation of the $n=1$ harmonic normalized with respect to the contribution
 of the 3-rd order Kerr nonlinearity.
Results are shown for two assumed bandgap energies of 7.5 and 8 \si{\electronvolt} and correspond to parallel polarization of pump and probe fields. {\bf (b)} Intensity dependence of the effective order of nonlinearity extracted from the simulated $n=1$ harmonic yield ratio for parallel and perpendicular polarization of pump and probe fields. The data compares results for the same values of the bandgap energy as in (a). Note that the right scale displays the associated intensity ratio. }
\label{fig:num1}
\end{figure}

Besides the identification of the leading role of ionization-induced nonlinearities, a comparison of the contributions from the first and third terms in Eq.~(\ref{eq:harm}) reveals that the harmonic emission arising from the acceleration term considered by Brunel is substantially weaker than the signal from the injection current (see Fig.~\ref{fig:num1}a) for a wide intensity range. The dominance of the signal resulting from the injection current and the general shape of the curve describing the intensity dependence of the effective order of nonlinearity are robust against changes of the ionization rate $\Gamma$ due to different assumed values of the bandgap, see blue and orange lines in Figs.~\ref{fig:num1}a and \ref{fig:num1}b. The intensity at which the rapid growth of the effective order occurs, however, clearly increases with the bandgap energy. This justifies our expectation that the injection-current-induced nonlinearity will be dominant also in other materials and indicates the potential to employ the wave-mixing signals for the time-resolved optical monitoring of bandgap modifications.

The main result of our study is the identification of the injection current as the source of a so far essentially overlooked but dominant nonlinearity for low-order wave mixing in solid dielectrics under the influence of strong fields.  The fact that the associated nonlinear polarization exceeds the magnitude of the low-order Kerr-response makes the inclusion of the injection current imperative in all simulations of nonlinear pulse propagation for field intensities near the damage threshold. Because of its sensitivity to the evolution of the ionization rate with field intensity, measurements and calculations of the polarization dependence of two-color wave mixing and its analysis in terms of the effective nonlinearity marks a new route for the time-resolved characterization of ionization dynamics in dielectrics, possibly even on ultrashort timescales down to the single cycle limit. Such knowledge would be of major importance for the analysis and optimization of strong-field-induced material modifications as it may enable the distinction of laser-driven strong-field ionization from other processes such as electron-impact ionization avalanching in the laser nano- and microprocessing of dielectrics.
\bibliography{lohg}

\begin{addendum}
 \item We thank Dr. Marco Jup\'e (Laser Zentrum Hannover e. V.) for measuring the bandgap of the samples used in this study.
 \item[Competing Interests] The authors declare that they have no
competing financial interests.
 \item[Correspondence] Correspondence and requests for materials
should be addressed to T. Fennel~(email: fennel@uni-rostock.de) or A.~Mermillod-Blondin~(email: mermillod@mbi-berlin.de).
\end{addendum}

\end{document}